\documentclass{llncs}
\usepackage{graphicx}
\usepackage{url}
\usepackage{caption}

\usepackage{algorithm}
\usepackage[noend]{algorithmic}

\usepackage{color}

\graphicspath{{images//}} 

\begin{document}
\frontmatter          
\pagestyle{headings}  
%
\title{Towards Linear Time Overlapping Community Detection in Social Networks}
\titlerunning{SLPA}  
%
\author{Jierui Xie \and Boleslaw K. Szymanski}

%

\institute{Rensselaer Polytechnic Institute\\ Troy, New York 12180, USA\\
\email{\{xiej2,szymansk\}@cs.rpi.edu}
}

\maketitle              

\begin{abstract}
Membership diversity is a characteristic aspect of social networks in which a person may belong to more than one social group. For this reason, discovering \textit{overlapping} structures is necessary for realistic social analysis. In this paper, we present a fast 
algorithm\footnote{Available at \url{https://sites.google.com/site/communitydetectionslpa/}},  called SLPA,
 for overlapping community detection in large-scale networks. SLPA spreads labels according to dynamic interaction rules. It
can be applied to both \textit{unipartite} and \textit{bipartite} networks. It is also able to uncover overlapping \textit{nested hierarchy}.
The time complexity of SLPA scales \textit{linearly} with the number of edges in the network.
Experiments in both synthetic and real-world networks show that SLPA has an excellent performance in identifying both  \textit{node} and \textit{community} level overlapping structures. 
\end{abstract}
\section{Introduction}
\label{sec:intr}
Community or modular structure is considered to be a significant property of real-world social networks. Thus, numerous techniques have been developed for effective community detection. However, most of the work has been done on \textit{disjoint} community detection. It has been well understood that people in a real social network are naturally characterized by \textit{multiple} community memberships. For example, a person usually has connections to several social groups like family, friends and colleges; a researcher may be active in several areas; in the Internet, a person can simultaneously subscribe to an arbitrary number of groups. 

For this reason, overlapping community detection algorithms have been investigated. These algorithms aim to discover a \textit{cover} \cite{LFM:2009}, defined as a set of clusters in which each node belongs to at least one cluster. In this paper, we propose an efficient algorithm for detecting both individual overlapping nodes and overlapping communities using the underlying network structure alone.

\section{Related Work}
\label{sec:rw}
We review the state of the art and categorize existing algorithms into five classes that reflect how communities are  identified.

\textbf{Clique Percolation}: CPM \cite{CPM:2005} is based on the assumption that a community consists of fully connected subgraphs and detects overlapping communities by searching for \textit{adjacent} cliques. 
CPMw \cite{CPMwFarkas:2007} extends CPM for weighted networks by introducing a subgraph intensity threshold.  

\textbf{Local Expansion}: The iterative scan algorithm (IS) \cite{Baumes2:2005},\cite{stephenRPI:2009} expands small cluster cores by adding or removing nodes until a local density function cannot be improved. The quality of seeds dictates the quality of discovered communities. LFM \cite{LFM:2009} expands a community from a random node. The size and quality of the detected communities depends significantly on the resolution parameter of the fitness function. EAGLE \cite{EAGLE:2009} and GCE \cite{GCE:2010} start with all maximal cliques in the network as initial communities. EAGLE uses the agglomerative framework to produce a dendrogram in $O(n^2s)$ time, where $n$ is the number of nodes, and $s$ is the maximal number of join operations. In GCE communities that are similar within a distance $\epsilon$ are removed. The greedy expansion takes $O(mh)$ time, where $m$ is the number of edges, and $h$ is the number of cliques.

\textbf{Fuzzy Clustering}: Zhang \cite{ZhangFuzzy:2007} used the spectral method to embed the graph into low dimensionality Euclidean space. Nodes are then clustered by the fuzzy c-mean algorithm. Psorakis et al. \cite{BayesianNMFIoannis:2011} proposed a model based on Bayesian nonnegative matrix factorization (NMF). These algorithms need to determine the number of communities $K$ and the use of matrix multiplication makes them inefficient. For NMF, the complexity is $O(Kn^2)$. 

\textbf{Link Partitioning}: Partitioning links instead of nodes to discover communities has been explored, where the node partition of a link graph leads to an edge partition of the original graph. In \cite{YYLinkClustering:2010}, single-linkage hierarchical clustering is used to build a link dendrogram. The time complexity is $O(nk_{max}^2)$, where $k_{max}$ is the highest degree of the $n$ nodes.

\textbf{Dynamical Algorithms}: Label propagation algorithms such as \cite{Raghavan:2007,Gregory:2010,JieruiXieLPA:2010} use labels to uncover communities. In COPRA \cite{Gregory:2010},  each node updates its belonging coefficients by \textit{averaging} the coefficients from all its neighbors in a synchronous fashion. The time complexity is $O(vm \log (vm/n))$ per iteration, where parameter $v$ controls the maximum number of communities with which a node can associate, $m$ and $n$ are the number of edges and number of nodes respectively. 

\section{SLPA: Speaker-listener Label Propagation Algorithm}
\label{sec:SLPA}
Our algorithm is an extension of the Label Propagation Algorithm (LPA) \cite{Raghavan:2007}. In LPA, each node holds only a single label and iteratively updates it to its neighborhood majority label. Disjoint communities are discovered when the algorithm converges. Like \cite{Gregory:2010}, our algorithm accounts for \textit{overlap} by allowing each node to possess multiple labels but it uses different dynamics with more general features.

SLPA mimics human pairwise communication behavior. In each communication step, one node is a speaker (information provider), and the other is a listener (information consumer). Unlike other algorithms,  each node has a \textit{memory} of the labels received in the past and takes its content into account to make the current decisions. This allows SLPA to avoid producing a number of small size communities as opposed to other algorithms. In a nutshell, SLPA consists of the following three stages:
\begin{algorithm}
\caption[caption]{: SLPA(\textit{T}, \textit{r})}
\label{alg1}
\begin{algorithmic}
\STATE $T$: the user defined maximum iteration 
\STATE $r$: post-processing threshold
\STATE 1) First, the \textit{memory} of each node is initialized with a unique label.
\STATE 2) Then, the following steps are repeated until the maximum iteration $T$ is reached:
\STATE 		\hspace{0.5cm} a. One node is selected as a listener.
\STATE 		\hspace{0.5cm} b. Each neighbor of the selected node randomly selects a label with probability \\ 
 					\hspace{0.8cm} proportional to the occurrence frequency of this label in its memory and sends \\
 					\hspace{0.8cm} the selected label to the listener.
\STATE 		\hspace{0.5cm} c. The listener adds the most popular label received to its memory.
\STATE 3) Finally, the post-processing based on the labels in the memories and the threshold\\
				  \hspace{0.4cm} $r$ is applied to output the communities.
\end{algorithmic}
\end{algorithm}	

Note that SLPA starts with each node being in its own community (a total of $n$), the algorithm explores the network and outputs the desired number of communities in the end. As such, the number of communities is not required as an input. Due to the step $c$, the size of memory increases by one for each node at each step. SLPA reduces to LPA when the size of memory is limited to one and the stop criterion is convergence of all labels. Empirically, SLPA produces relatively stable outputs, independent of network size or structure, when \textit{T} is greater than 20. Although SLPA is non-deterministic due to the random selection and ties, it performs well on average as shown in later sections.   

\begin{figure}[t]
\centering	 
	 \begin{minipage}[t]{0.45\linewidth}
	 \centering	
	 	\includegraphics[scale=0.6]{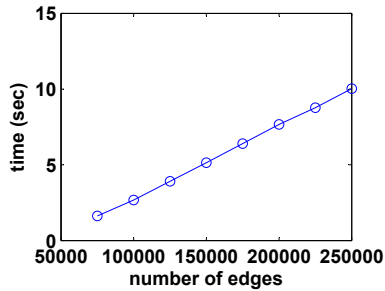}
			\caption{The execution times of SLPA in synthetic networks with $n=5000$ and average degree $\overline{k}$ varying from 10 to 80.}
			\label{fig:executionTime}	 
			\end{minipage}
\end{figure}	
		 	
\textbf{Post-processing and Community Detection}: In SLPA, the detection of communities is performed when the stored information is post-processed. Given the memory of a node, SLPA converts it into a probability distribution of labels. Since labels represent community id's, this distribution naturally defines the \textit{strength} of association to communities to which the node belongs. To produce \textit{crisp} communities in which the membership of a node to a given community is \textit{binary}, i.e., either a node is in a community or not, a simple thresholding procedure is performed: if the probability of seeing a particular label during the whole process is less than a given threshold $r\in [0,0.5]$, this label is deleted. After thresholding, connected nodes having a particular label are grouped together and form a community. If a node contains multiple labels, it belongs to more than one community and is called an \textit{overlapping node}. A smaller value of $r$ produces a larger number of communities. However, the effective range is typically narrow in practice.
When $r\geq 0.5$, SLPA outputs disjoint communities.

\textbf{Complexity}: The initialization of labels requires $O(n)$, where $n$ is the total number of nodes. The outer loop is controlled by the user defined maximum iteration $\textit{T}$, which is a small constant which in our experiments was set to 100. The inner loop is controlled by $n$. Each operation of the inner loop executes one speaking rule and one listening rule. The speaking rule requires exactly $O(1)$ operation. The listening rule takes $O(\overline{k})$ on average, where $\overline{k}$ is the average node degree. In the post-processing, the thresholding operation requires $O(Tn)$ operations since each node has a memory of size $\textit{T}$. In summary, the time complexity of the entire algorithm is $O(Tn\overline{k})$ or $O(Tm)$, linear with the total number of edges $m$. 
The execution times for synthetic networks where averaged for each $\overline{k}$ 
over networks with different structures, i.e., different degree and community size distributions (see Section \ref{sec:Methodology} for details).
The results shown in Fig. \ref{fig:executionTime} confirm the \textit{linear} 
scaling of the execution times. 
On a desktop with 2.80GHz CPU, SLPA took about six minutes to run over 
a two million nodes Amazon co-purchasing, which is ten times faster than GCE 
running over the same network.
\section{Tests in Synthetic Networks}
\label{sec:ExpLFR}
\subsection{Methodology} 
\label{sec:Methodology}
To study the behavior of SLPA, we conducted extensive experiments in both synthetic and real-world networks. For synthetic random networks, we adopted the widely used LFR benchmark\footnote{\url{http://sites.google.com/site/andrealancichinetti/files}} \cite{LFR:2008}, which allows heterogeneous distributions of node degrees and community sizes.

\begin{minipage}[t]{.45\textwidth} 
  \centering
  \captionof{table}{Algorithms in the tests.}
  \label{table:algs}
  \scalebox{0.8} {
	\small\addtolength{\tabcolsep}{-1pt}  
		\begin{tabular}{cccc} \hline
		\textbf{Algorithm} & \textbf{Complexity} & \textbf{Imp} \\ \hline
 			CFinder \cite{CPM:2005}, 2005 & - & C++ \\ 
 			LFM \cite{LFM:2009}, 2009 & $O(n^2)$ & C++ \\ 
 			EAGLE \cite{EAGLE:2009}, 2009 & $O(n^2s)$ & C++ \\ 
 			CIS \cite{stephenRPI:2009}, 2009 &  $O(n^2)$ & C++ \\ 
 			GCE \cite{GCE:2010}, 2010 & $O(mh)$ & C++ \\ 
 			COPRA \cite{Gregory:2010}, 2010 & $O(vm \log (vm/n))$ & Java \\ 
 			NMF \cite{BayesianNMFIoannis:2011}, 2010 & $O(Kn^2)$ & Matlab \\ 
 			Link \cite{YYLinkClustering:2010}, 2010 & $O(nk_{max}^2)$ & C++ \\ 
 			SLPA, 2011 & $O(Tm)$ & C++ \\ \hline 
		\end{tabular}
		}
\end{minipage}\qquad
\begin{minipage}[t]{.45\textwidth}
  \centering
  \captionof{table}{The ranking of algorithms.}
  \label{table:ranking}
	\scalebox{0.8} {
	\small\addtolength{\tabcolsep}{-0pt}  
			\begin{tabular}{ccccccc} \hline
			\textbf{Rank} & \textbf{$RS_{Omega}$} &  \textbf{$RS_{NMI}$} & \textbf{$RS_{F}$} \\ \hline
			1 & SLPA     & SLPA     & SLPA     \\ 
			2 & COPRA    & GCE      & CFinder     \\ 
			3 & GCE      & NMF    & COPRA   \\ 
			4 & CIS       & CIS      & Link    \\ 
			5 & NMF      & LFM    &  LFM    \\ 
			6 & LFM      & COPRA     & CIS    \\ 
			7 & CFinder  & CFinder   & GCE      \\ 
			8 & Link    &  EAGLE &  EAGLE    \\ 
			9 & EAGLE   &  Link  & NMF      \\ \hline
		\end{tabular}
		}
\end{minipage}

We used networks with size $n=5000$. The average degree is kept at $\overline{k}=10$. The degree of overlapping is determined by two parameters. {\boldmath $O_n$} defines the number of overlapping nodes and is set to 10\% of all nodes. {\boldmath $O_m$} defines the number of communities to which each overlapping node belongs and varies from 2 to 8 indicating the diversity of overlap. By increasing the value of $O_m$, we create harder detection tasks. Other parameters are as follows: node degrees and community sizes are governed by the power laws with exponents 2 and 1; the maximum degree is 50; the community size varies from 20 to 100; the expected fraction of links of a node connecting it to other communities, called the mixing parameter $\mu$, is set to 0.3. We generated ten instance networks for each setting.
 
In Table \ref{table:algs}, we compared SLPA with eight other algorithms representing different categories discussed in section \ref{sec:rw}. For algorithms with tunable parameters, the performance with the optimal parameter is reported. For CFinder, \textit{k} varies from 3 to 10; for COPRA, \textit{v} varies from 1 to 10; for LFM $\alpha$ is set to 1.0 \cite{LFM:2009}. For $Link$, the threshold varies from 0.1 to 0.9 with an interval 0.1. For SLPA, the number of iterations \textit{T} is set to 100 and \textit{r} varies from 0.01 to 0.1. The average performance together with error bar over ten repetitions are reported for SLPA and COPRA. For NMF, we applied a threshold varying from 0.05 to 0.5 with an interval 0.05 to convert it to a crisp clustering. 

To summarize the vast amount of comparison results and provide a measure of relative performance, we proposed $RS_M(i)$, the averaged ranking for algorithm $i$ with respect to measure $M$ as follows:
  \begin{equation}
\label{eq:ranking}
		RS_M(i)=\sum_{j=1} w_j \cdot rank(i,O_m^j) ,
\end{equation}
 where $O_m^j$ is the number of memberships in $\{2,3,\cdots,8\}$, $w_j$ is the weight, and function \textit{rank} returns the ranking of algorithm $i$ for the given $O_m$. For simplicity, we assume equal weights over different $O_m$'s in this paper. Sorting $RS_M$ in increasing order gives the final ranking among algorithms.
\begin{figure}[t]	
\centering
	\begin{minipage}[t]{0.45\linewidth}
	\centering
    \includegraphics[scale=0.5]{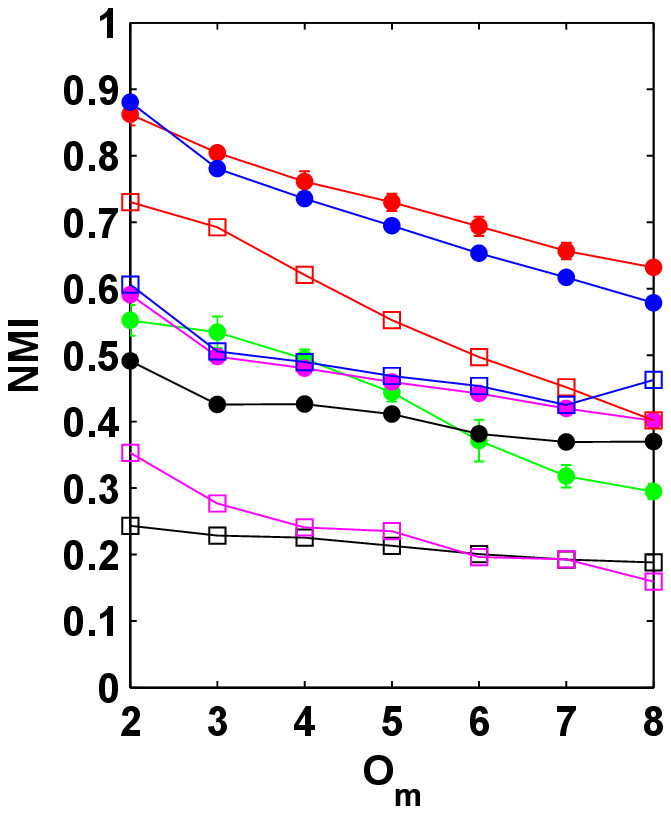} 
		 \caption{NMI as a function of the number of memberships $O_m$ in LFR.}
	 	 \label{fig:pakddNMILFRN5000mu03}
	\end{minipage}	
	 \hspace{0.1cm}  
	 \begin{minipage}[t]{0.45\linewidth}
	 \centering
     \includegraphics[scale=0.5]{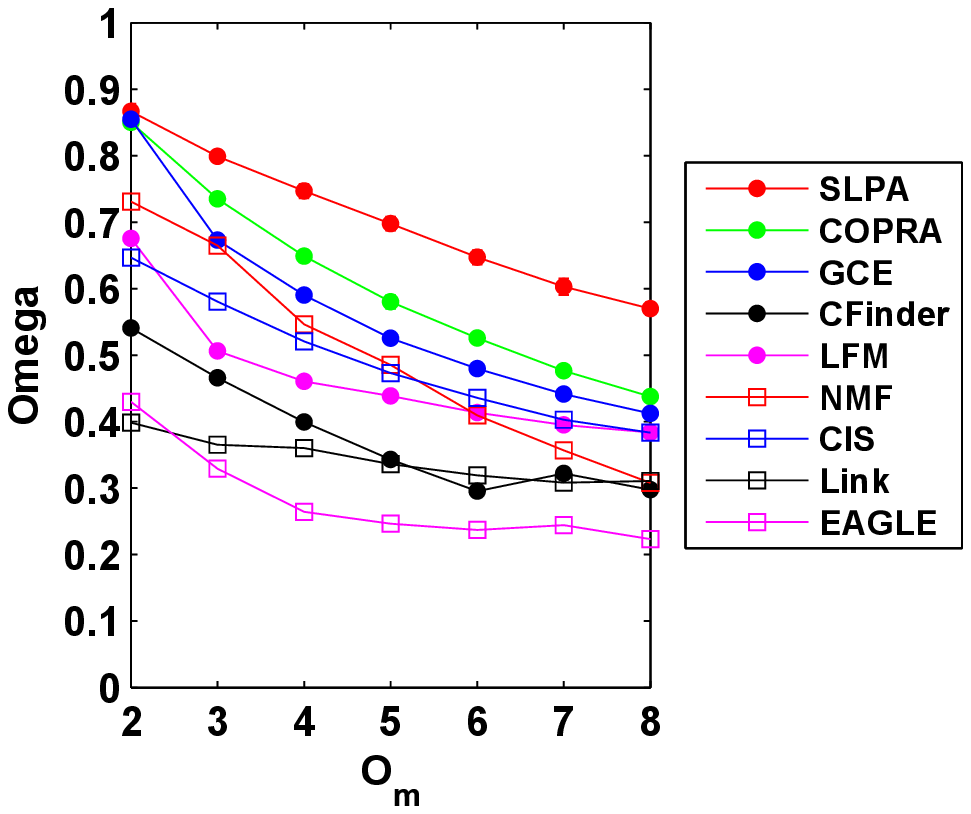} 
		 \caption{Omega as a function of the number of memberships $O_m$ in LFR.}
	 	 \label{fig:pakddOMGLFRN5000mu03}
	\end{minipage}	
\end{figure}
\begin{figure*}[t]	
\centering
\begin{minipage}[t]{0.45\linewidth}	
	\centering
		\includegraphics[scale=0.45]{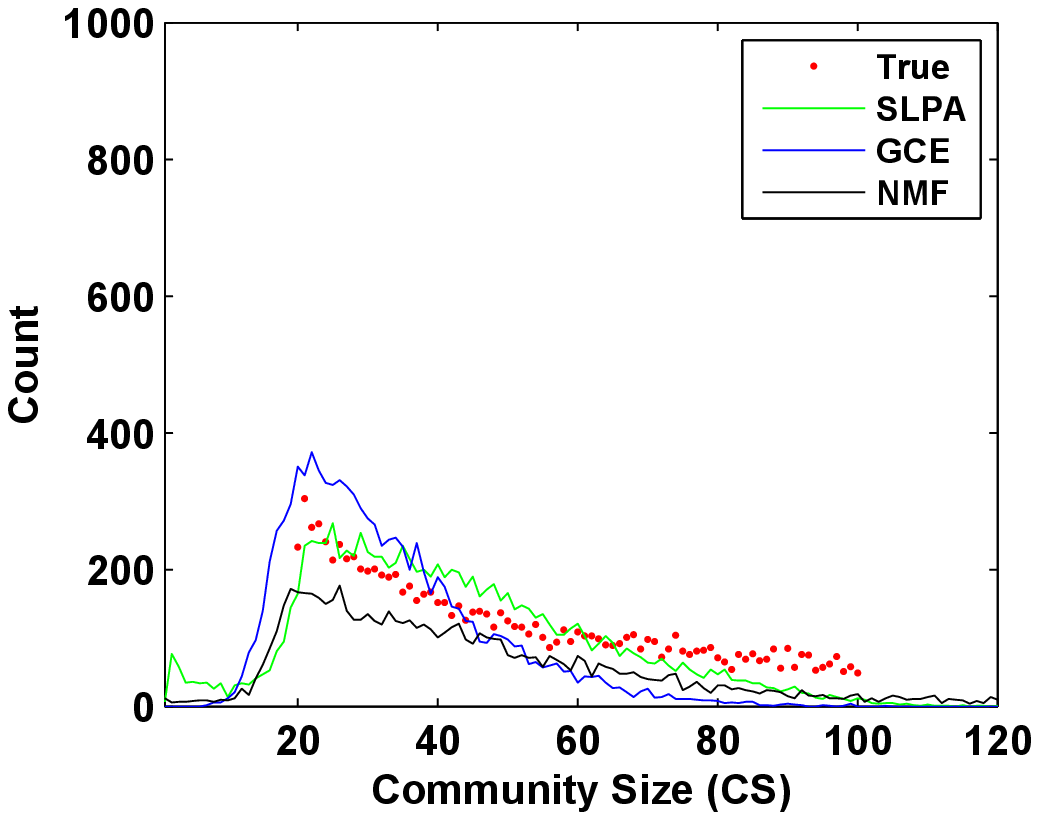} 
		\caption{The unimodal histogram of the detected community sizes for SLPA, GCE and NMF. }
	 	\label{fig:pakddcomdis1}				
	\end{minipage}	
	 \hspace{0.1cm}  
	 \begin{minipage}[t]{0.45\linewidth}
	 \centering
\includegraphics[scale=0.45]{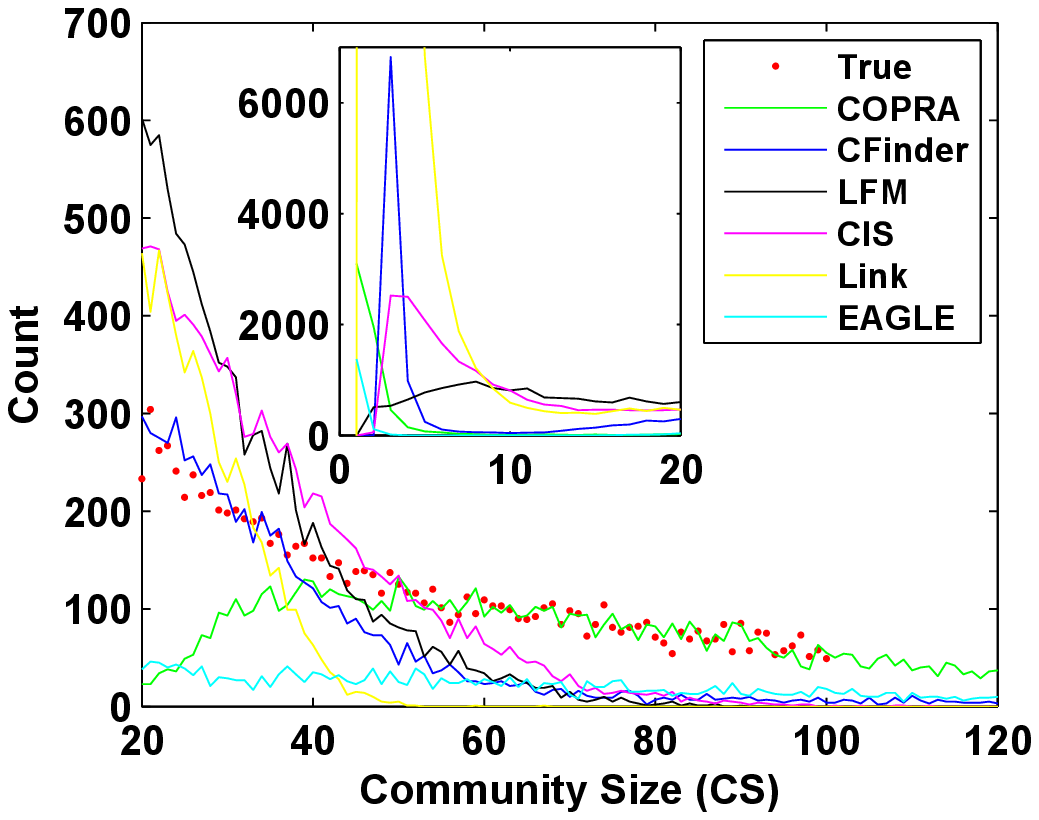} 
		\caption{The bimodal histogram of the detected community sizes  for COPRA, CIS, LFM, Link, EAGLE and CFinder.}
	 	\label{fig:pakddcomdis2}	
	\end{minipage}	
\end{figure*}
\subsection{Identifying Overlapping Communities in LFR}
The extended normalized mutual information (NMI) \cite{LFM:2009} and Omega Index \cite{Omega:1988} are used to quantify the quality of communities discovered by an algorithm.
NMI measures the fraction of nodes in agreement in two covers, while Omega is based on pairs of nodes. Both NMI and Omega yield the values between 0 and 1. The closer this value is to 1, the better the performance is.

As shown in Fig. \ref{fig:pakddNMILFRN5000mu03} and Fig. \ref{fig:pakddOMGLFRN5000mu03}, some algorithms behave differently under different measures (the rankings of $RS_{Omega}$ and $RS_{NMI}$ among algorithms in Table \ref{table:ranking} also change). As opposed to NMF and COPRA, which are especially sensitive to the measure, SLPA is remarkably stable in NMI and Omega. 

Comparing the detected and known numbers of communities and distributions of community sizes (CS) helps to understand the results. On one hand, we expect the community size to follow a power law with exponent 1 and to range from 20 to 100 by design. As shown in Fig. \ref{fig:pakddcomdis1}, high-ranking (with high NMI) algorithms such as SLPA, GCE and NMF typically yield a \textit{unimodal} distribution with a peak at $CS=20$  fitting well with the ground truth distribution. In contrast, algorithms in Fig. \ref{fig:pakddcomdis2} typically produce a \textit{bimodal} distribution. The existence of an extra dominant mode for CS ranging from 1 to 5 results in a significant number of small size communities in CFinder, LFM, COPRA, Link and CIS. These observations nicely explain the ranking with respect to NMI. 
\subsection{Identifying Overlapping Nodes in LFR}
\begin{figure*}[t]	
\centering
	\begin{minipage}[t]{0.45\linewidth}
	\centering
    \includegraphics[scale=0.5]{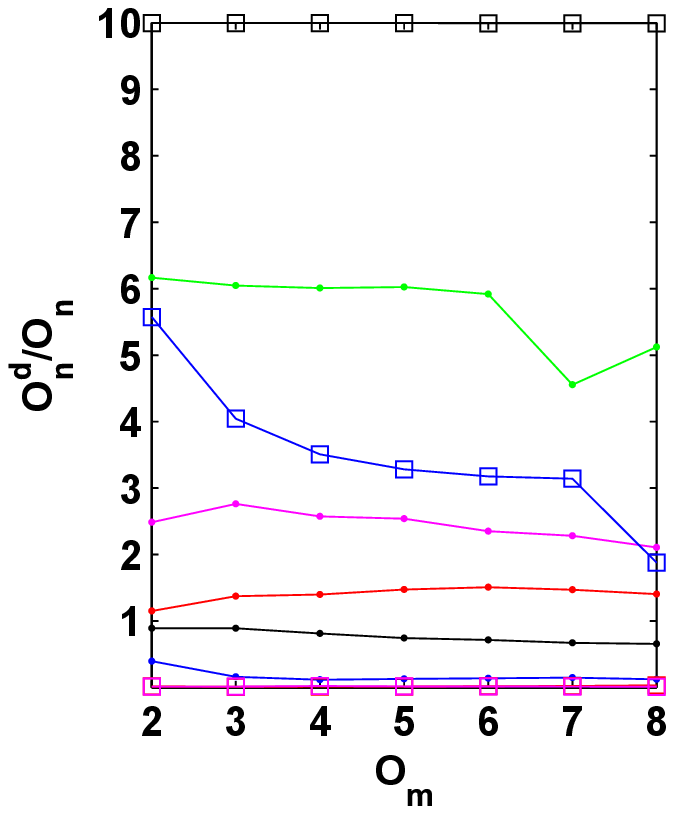} 
		\caption{The number of detected overlapping nodes relative to the
ground truth. }
	 	\label{fig:pakddnumDetOn}
	\end{minipage}	
	 \hspace{0.2cm}  
	 \begin{minipage}[t]{0.45\linewidth}
	 \centering
    \includegraphics[scale=0.5]{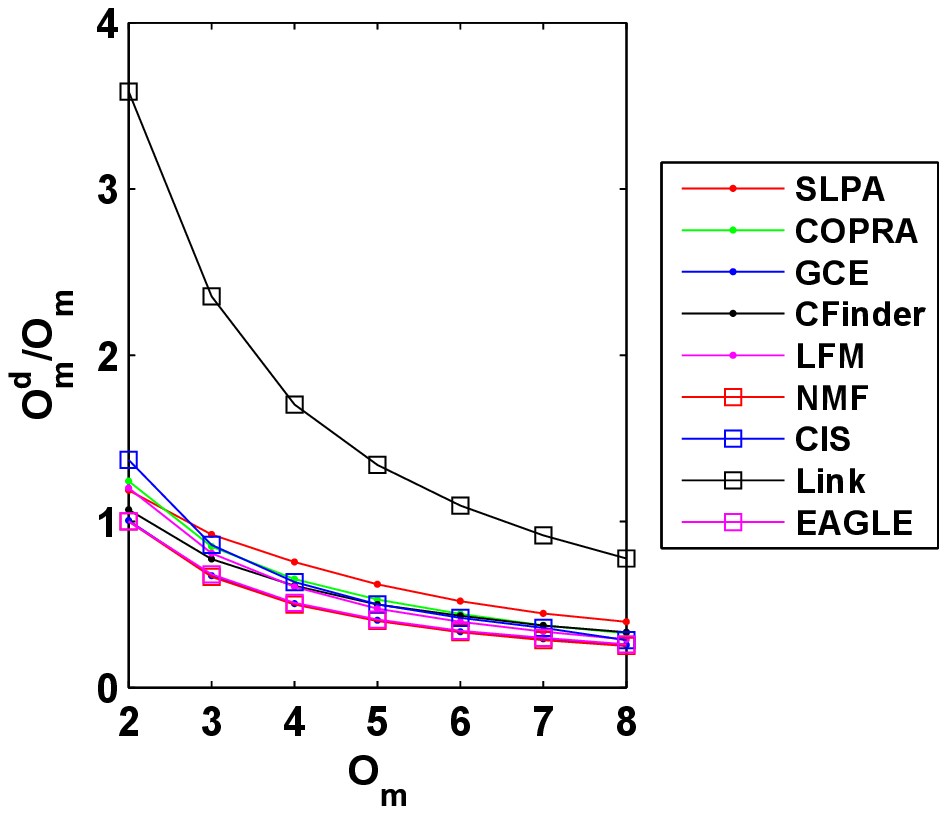} 
		\caption{The number of memberships of detected overlapping nodes relative to the
ground truth. }
		\vspace{0.3cm}
	 	\label{fig:pakddnumDetOm}
	\end{minipage}	
\end{figure*}

Identifying nodes overlapping multiple communities is an essential component of measuring the quality of a detection algorithm. However, the node level evaluation was often neglected. Here we first look at the number of detected overlapping nodes $O_n^d$ (see Fig. \ref{fig:pakddnumDetOn}) and detected memberships $O_m^d$ (see Fig. \ref{fig:pakddnumDetOm}) relative to the ground truth $O_n$ and $O_m$,
based on the information in Fig. \ref{fig:pakddNMILFRN5000mu03}. Note that a value close to 1 indicates
closeness to the ground truth, and values over 1 are possible when an algorithm detects more nodes or memberships than there are known to exist. As shown, SLPA yields the numbers that are close to the ground truth in both cases.

Note that $O_n^d$ alone is insufficient to accurately quantify the detection performance, as it contains both true and false positive. To provide precise analysis, we consider the identification of overlapping nodes as a \textit{binary classification} problem. A node is labeled as \textit{overlapping} as long as $O_m$\textgreater$1$ or $O_m^d$\textgreater$1$ and labeled as \textit{non-overlapping} otherwise. Within this framework, we can use F-score as a measure of detection accuracy defined as
\begin{equation}
\label{eq:F}
		F=\frac{2\cdot precision \cdot recall}{precision + recall} ,
\end{equation}
where \textit{recall} is the number of correctly detected overlapping nodes divided by $O_n$, and \textit{precision} is the number of correctly detected overlapping nodes divided by $O_n^d$. F-score reaches its best value at 1 and worst score at 0. 

\begin{figure}[t]	
\centering
	 \begin{minipage}[t]{0.31\linewidth}
	 \centering
    \includegraphics[scale=0.5]{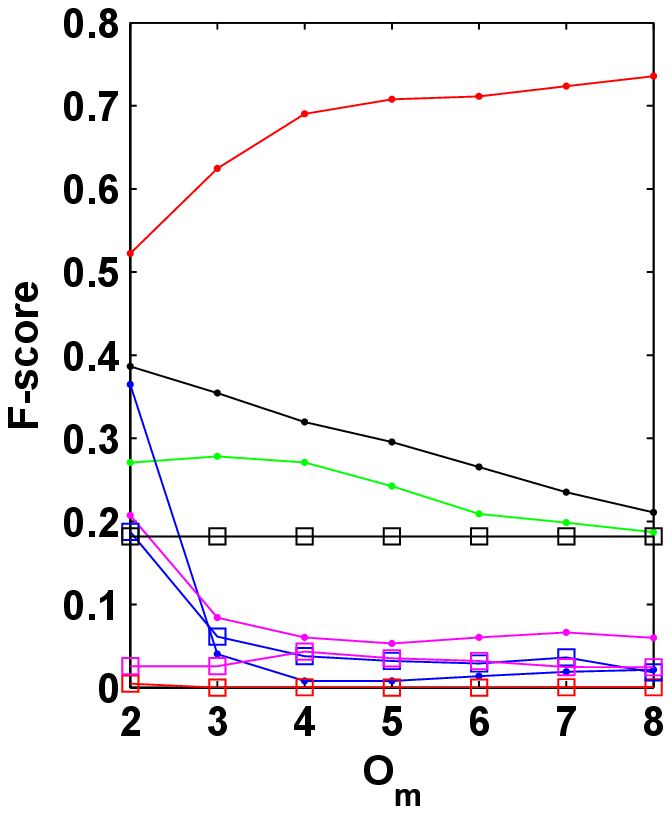} 
		\caption{The F-score.}
	 	\label{fig:pakddfscore}
	\end{minipage}	
	 \hspace{0.1cm}  
	\begin{minipage}[t]{0.31\linewidth}
	\centering
    \includegraphics[scale=0.5]{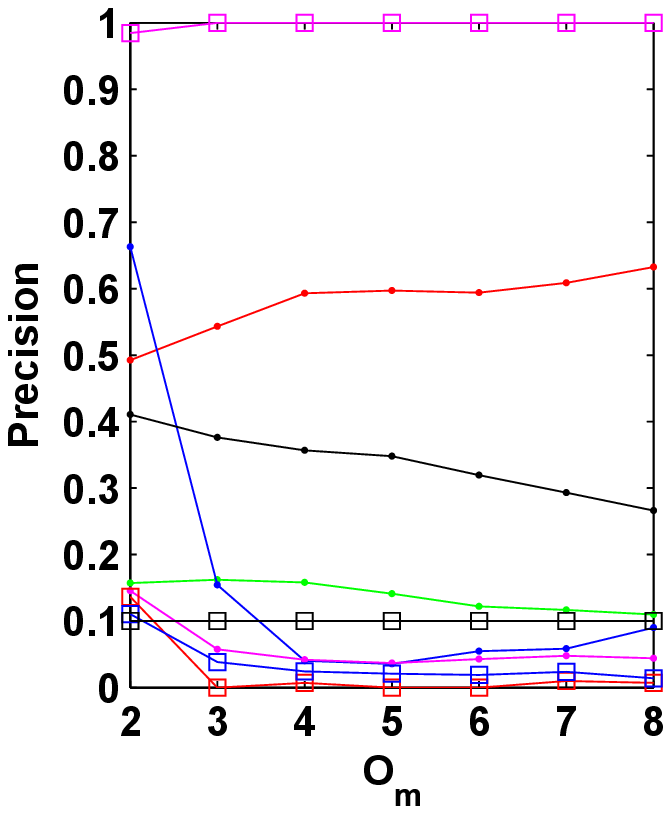} 
		\caption{The precision.}
	 	\label{fig:pakddprecision}
	\end{minipage}	
	 \hspace{0.1cm}  
	 \begin{minipage}[t]{0.31\linewidth}
	 \centering
    \includegraphics[scale=0.5]{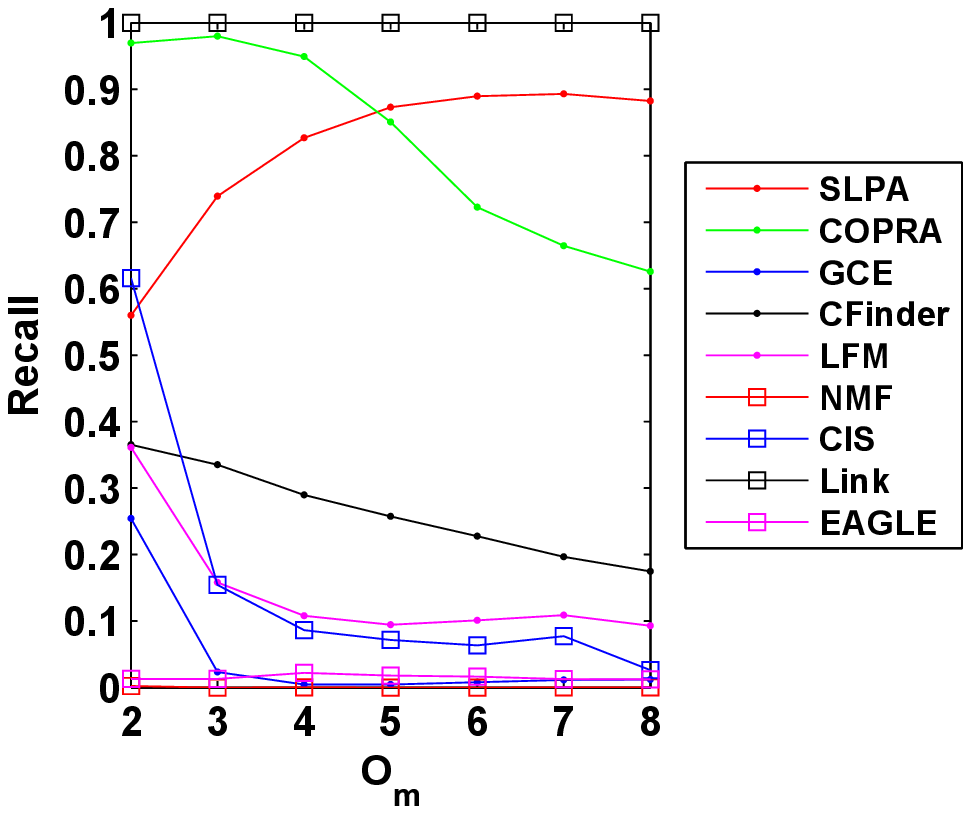} 
		\caption{The recall.}
		\vspace{0.3cm}
	 	\label{fig:pakddrecall}
	\end{minipage}	
\end{figure}

As shown in Fig. \ref{fig:pakddfscore}, SLPA achieves the best score on this metric. This score has a positive correlation with $O_m$ while scores of other algorithms are negatively correlated with it. SLPA correctly uncovers a reasonable fraction of overlapping nodes even when those nodes belong to many groups (as demonstrated by the high precision and recall in Fig. \ref{fig:pakddprecision} and Fig. \ref{fig:pakddrecall}). Other algorithms that fail to have a good balance between precision and recall result in low F-score, especially for EAGLE and Link. The high precision of EAGLE (also CFinder and GCE for $O_m=2$) shows that clique-like assumption of communities may help to identify overlapping nodes. However, they under-detect the number of such nodes. 

With the F-score ranking, GCE and NMF no longer rank in the top three algorithms, while SLPA stays there. Taking both community level performance (NMI and Omega) and node level performance (F-score) into account, we conclude that SLPA performs well in the LFR benchmarks.
%
\section{Tests in Real-world Social Networks}
We applied SLPA to a wide range of well-known social networks\footnote{ \url{www-personal.umich.edu/~mejn/netdata/} and  \url{snap.stanford.edu/data/}} as listed in Table \ref{tab:socailNets}. The high school friendship networks that were analyzed in a project funded by the National Institute of Child Health and Human Development, are social networks in high schools self-reported by students together with their grades, races and sexes. We used these additional attributes for verification.

\begin{table}[t]
\centering
\caption{Social networks in the tests}
\label{tab:socailNets}
  \scalebox{0.9}{ 
  \begin{tabular}{cccccc} \hline
			\textbf{Network} & {\boldmath $n$} &  \textbf{$\overline{k}$} & \textbf{Network} & {\boldmath $n$} &  \textbf{$\overline{k}$} \\ \hline
			 karate (KR)   	& 34  	& 4.5 & Email (EM)	 		  & 33696 & 10.7\\ 
			 football (FB) 	& 115   & 10.6 & P2P	 	 	 	 		& 62561 & 2.4  \\ 
			 lesmis (LS)   	& 77   	& 6.6 & Epinions (EP)		& 75877 & 10.6 \\ 
			 dolphins (DP)		& 62  	& 5.1 & Amazon (AM)	 		& 262111& 6.8 \\ 
			 CA-GrQc (CA)		& 4730  & 5.6 & HighSchool (HS1) & 69  	& 6.3  \\ 
			 PGP	 	 	 	  	& 10680 & 4.5 & HighSchool (HS2) & 612  	& 8.0 \\ \hline 
		\end{tabular}
		}
		\vspace{0.35cm}
\end{table}
\begin{figure*}[t]	
\begin{minipage}[t]{0.55\linewidth}
   \includegraphics[scale=0.55]{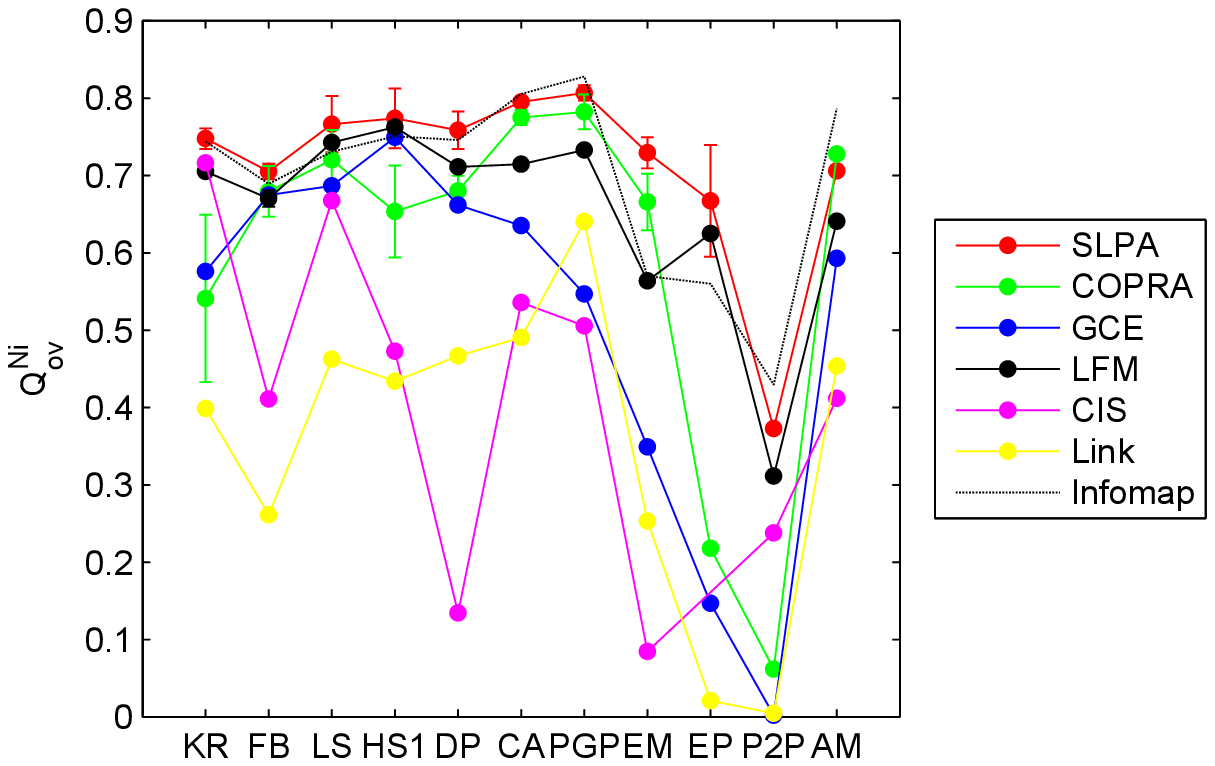} 
		\caption{Overlapping modularity $Q_{ov}^{Ni}$.}
	 	\label{fig:pakddRealQovNi} 
	\end{minipage}	
	 \hspace{0.1cm}  
	 \begin{minipage}[t]{0.45\linewidth}
	 \centering
     \includegraphics[scale=0.5]{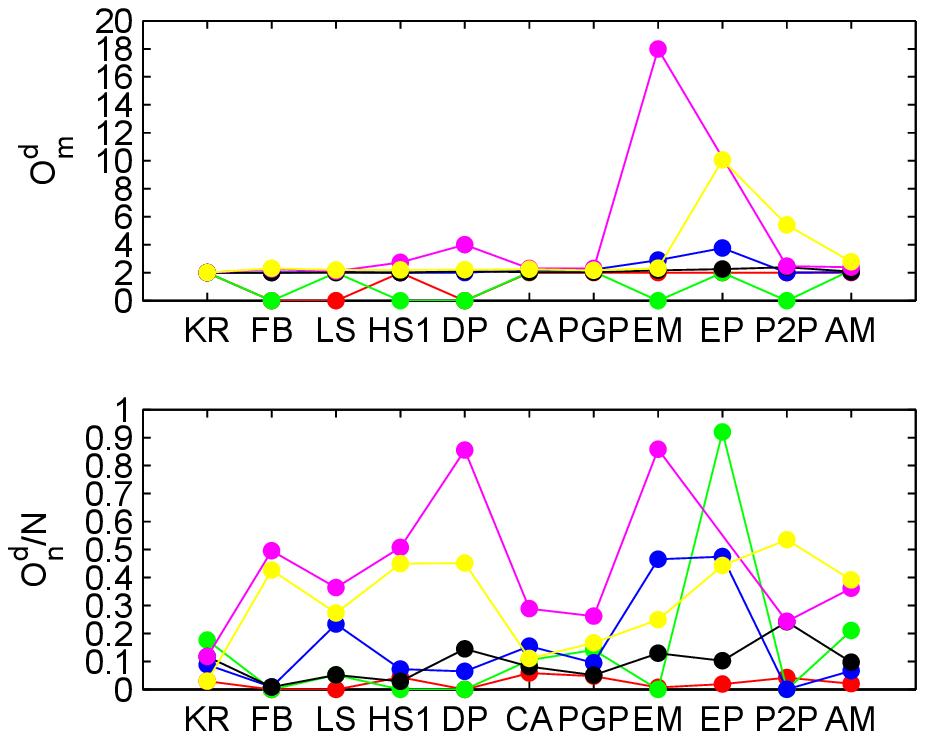} 
		 	\caption{The number of detected memberships (top) and the fraction of detected overlapping nodes (bottom).}
	 	\label{fig:pakddRealQovNi-OmOn}
	\end{minipage}	
\end{figure*}
\subsection{Identifying Overlapping Communities in Social Networks} 
To quantify the performance, we used the overlapping modularity, $Q_{ov}^{Ni}$ (with values between 0 and 1), proposed by Nicosia \cite{Nicosia:2009}, which is an extension of Newman's modularity. A high value indicates a significant overlapping community structure relative to the null model. We removed CFinder, EAGLE and NMF from the test because of either their memory or their computation inefficiency on large networks. As an additional reference, we added the disjoint detection results with the Infomap algorithm. 

As shown in Fig. \ref{fig:pakddRealQovNi}, in general, SLPA achieves the highest average $Q_{ov}^{Ni}$, followed by LFM and COPRA, even though the performance of SLPA has larger fluctuation than that in synthetic networks. Compared with COPRA, SLPA is more stable as evidenced by smaller deviation of its $Q_{ov}^{Ni}$ score. In contrast, COPRA does not work well on highly sparse networks such as \textit{P2P}, for which COPRA finds merely one single giant community. COPRA also fails on \textit{Epinions} network because it claims too many overlapping nodes in view of consensus of other algorithms as seen in the bottom of Fig. \ref{fig:pakddRealQovNi-OmOn}. Such over-detection also applies to CIS and Link, resulting in low $Q_{ov}^{Ni}$ scores for these two algorithms. The results in Fig. \ref{fig:pakddRealQovNi-OmOn} (based on the clustering with the best $Q_{ov}^{Ni}$) show a common feature in the tested real-world networks, which is a relatively little agreement between results of different algorithms, i.e., the relatively small overlap in both the fraction of overlapping nodes (typically less than 30\%) and the number of communities of which an overlapping node is a member (typically 2 or 3). 

\begin{figure}[tp]	
  \centering
	\includegraphics[scale=0.5]{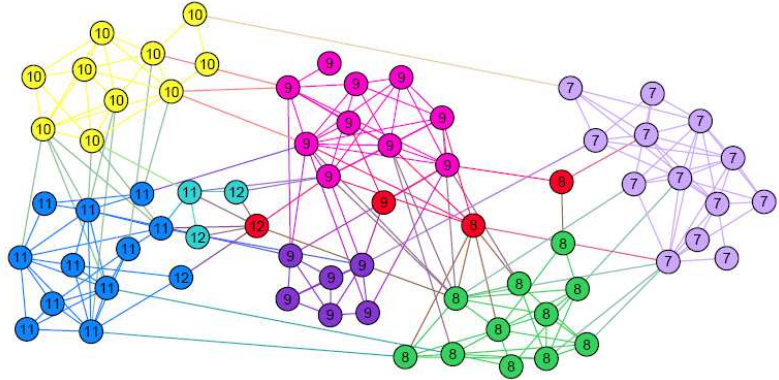}
	\caption{High school network ($n=69$, $\overline{k}=6.4$). Labels are the known grades ranging from 7 to 12. Colors represent communities discovered by SLPA. The overlapping nodes are highlighted by red color.}
	\vspace{0.2cm}
	\label{fig:comm1}
\end{figure}

As known, a high modularity might not necessarily result in a \textit{true} partitioning as it does in the disjoint community detection. We used the high school network (HS1) with known attributes to verify the output of SLPA. As shown in Fig. \ref{fig:comm1}, there is a good agreement between the found and known partitions in term of student's \textit{grades}. In SLPA, the grade 9 community is further divided into two subgroups. The larger group contains only white students, while the smaller group demonstrates \textit{race} diversity. These two groups are connected partially via an overlapping node. It is also clear that overlapping nodes only exist on the boundaries of communities. A few overlapping nodes are assigned to three communities, while the others are assigned to two communities (i.e., their $O_m$ is 2).

\subsection{Identifying Overlapping Communities in Bipartite Networks}
Discovering communities in bipartite networks is important because they provide a natural representation of many social networks. One example is the online tagging system with both users and tags. Unlike the original LPA algorithm, which performs poorly on bipartite networks, SLPA works well on this kind of networks. We demonstrate this using two real-world networks\footnote{Data are available at \url{http://toreopsahl.com/datasets/}}. One is a Facebook-like social network. One type of nodes represents users (abbr. FB-M1), while the other represents messages (abbr. FB-M2). The second network is the interlocking directorate. One type of nodes represents affiliations (abbr. IL-M1), while the other individuals (abbr. IL-M2).  

We compared SLPA with COPRA in Table \ref{table:bipartite}. One difference between SLPA and COPRA is that SLPA applies to the \textit{entire} bipartite network directly, while COPRA is applied to each type of nodes \textit{alternatively}. $Q_{ov}^{Ni}$ is computed on the \textit{projection} of each type of nodes. Again, we allow \textit{overlapping} between communities. Although COPRA is slightly better (by 0.03) than SLPA on the second type of nodes for interlock network, it is much worse (by 0.11) on the first type. Moreover, COPRA fails to detect meaningful communities in the Facebook-like network, while SLPA demonstrates relatively good performance. 
\begin{table}[hbpt]
\centering
\caption{The $Q_{ov}^{Ni}$ of SLPA and COPRA for two bipartite networks.}
\label{table:bipartite}
\small\addtolength{\tabcolsep}{-0pt}  
\scalebox{0.8}{
\begin{tabular}{cccccccc} \hline
\textbf{Network} & {\boldmath $n$}	& \textbf{SLPA (std)}	& \textbf{COPRA (std)}	\\ \hline 							
	FB-M1&	899&	0.23 (0.10) & 0.02 (0.07)\\ 
	FB-M2&	522&	0.36 (0.02)&	0.02 (0.07)\\ 
	IL-M1&	239&	0.59 (0.02)&	0.48 (0.02)\\ 
	IL-M2&	923&	0.69 (0.01)&	0.72 (0.01)\\ \hline
\end{tabular}
}
\end{table}
\subsection{Identifying Overlapping Nested Communities}
\begin{figure*}[t]
  \begin{minipage}[t]{1\linewidth}
		\centering
		\includegraphics[scale=0.8]{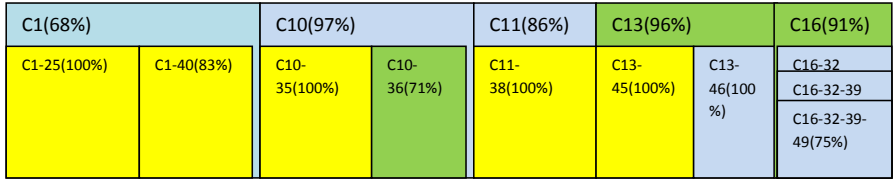}
		\caption{The nested structure in the high school network represented as a Treemap. The color represents the best explaining attribute: \textcolor{blue}{\textit{blue}} for \textit{grade}; \textcolor{green}{\textit{green}} for \textit{race}; and \textcolor{yellow}{\textit{yellow}} for \textit{sex}. Numbers in parenthesis are the matching scores defined in the text. The size of shapes is proportional to the community size. Due to the page limit, only part of the entire treemap is shown.}
		\label{fig:HStree}
	\end{minipage}	
\end{figure*}

In the above experiments, we applied a post-processing to remove subset communities from the raw output of stages 1 and 2 of SLPA. This may not be necessary for some applications. Here, we show that rich \textit{nested} structure can be recovered in the high school network (HS2) 
 with $n=612$. The hierarchy is shown as a treemap\footnote{Treemap is used for visualization: \url{www.cs.umd.edu/hcil/treemap/}.} shown in Fig. \ref{fig:HStree}. To evaluate the degree to which a discovered community matches the known attributes, we define a \textit{matching score} as the \textit{largest} fraction of matched nodes relative to the community size among three attributes (i.e., grade, race and sex). The corresponding attribute is said to best explain the community found by SLPA.

As shown, SLPA discovers a tree with a height of four. Most of the communities are distributed on the first two levels. The community name shows the full hierarchy path (connected by a dash '-') leading to this community. For example, \textit{C1} has id 1 and is located on the first level, while \textit{C1-25} has id 25, and it is the second level sub-community of community with id 1.

Nested structures are found across different attributes. For example, \textit{C13} is best explained by \textit{race}, while its two 
sub-communities perfectly account for \textit{grade} and \textit{sex} respectively. In \textit{C1}, sub-communities explained by the same attribute account for \textit{different} attribute \textit{values}. For example, both \textit{C1-25} and \textit{C1-40} are identified by \textit{sex}. However, the former contains only \textit{male} students, while in the latter \textit{female} students are the majority. Although the treemap is not capable of displaying overlaps between communities, the nested structures overlap as before.
 
\section{Conclusions}
We introduced a dynamic interaction process, SLPA as a basis for an efficient and effective \textit{unified} overlapping community detection algorithm. SLPA allows us to analyze different kinds of community structures, such as disjoint communities, individual overlapping nodes, overlapping communities and overlapping nested hierarchy in both unipartite and bipartite topologies. Its underlying process can be easily modified to accommodate other types of networks (e.g., k-partite graphs). In the future work, we plan to apply SLPA to temporal community detection.

\textbf{Acknowledgments.} Research was sponsored by the Army Research Laboratory and was accomplished
under Cooperative Agreement Number W911NF-09-2-0053 and by the Office of Naval 
Research Grant No. N00014-09-1-0607. The views and conclusions contained in this 
document are those of the authors and should not be interpreted as representing the 
official policies, either expressed or implied, of the Army Research Laboratory or the 
Office of Naval Research or the U.S. Government. The U.S. Government is authorized 
to reproduce and distribute reprints for Government purposes notwithstanding any 
copyright notation here on.

\bibliographystyle{splncs03}
\bibliography{CommunityBIB-Jerry}

\end{document}